# Comparative study on radiation resistance of WTaCrV high-entropy alloy and tungsten in helium-containing conditions

Amin Esfandiarpour[1*], Damian Kalita[1], Zbigniew Koziol[2], and Mikko Alava[1,3]

[1] NOMATEN Centre of Excellence, National Centre for Nuclear Research, Otwock, Andrzeja Soltana 7, 05-400, Poland

[2] National Center for Nuclear Research, Materials Research Laboratory, ul. Andrzeja Sołtana 7, Otwock-Świerk, 05-400, Poland

[3] Department of Applied Physics, Aalto University, P.O. Box 11000, 00076 Aalto, Espoo, Finland

* Corresponding author amin.esfandiarpour@ncbj.gov.pl (Amin Esfandiarpour)

**Abstract**

W and W-based high-entropy alloys (HEAs) are promising candidates for plasma-facing materials in fusion reactors. While irradiation studies on W have revealed a tendency for helium (He) bubble formation and radiation-induced defects, investigations of WTaCrV HEA have demonstrated superior radiation resistance, whether under $He^+$ irradiation or heavy ion irradiation. To assess material performance under conditions relevant to fusion reactors—characterized by fast neutrons and gas production from transmutation reactions—complex irradiation environments need to be modeled. Using molecular dynamics simulations, we examined defect evolution in W and equimolar WTaCrV HEA with and without preexisting He atoms under cascade overlap conditions up to 0.2 dpa at 300 K. In W, dislocation loops and large interstitial clusters formed readily, with increasing He content leading to higher dislocation densities and the formation of polygonal interstitial networks. In contrast, WTaCrV alloy exhibited strong resistance to formation of dislocation loop and large interstitial clusters but was more susceptible to formation of bubble at higher He concentrations. Bubble growth was driven by helium trapping at vacancy sites and the coalescence of smaller bubbles. Larger bubbles remained stable against cascade overlap, limiting further growth by coalescence.

**Introduction**

The development of advanced materials capable of withstanding the extreme conditions in nuclear fusion reactors represents a critical challenge for achieving sustainable fusion energy[1,2]. Plasma-facing materials (PFMs), which are directly exposed to intense particle fluxes and extreme heat from high-temperature plasma, must simultaneously resist high-energy neutron irradiation and withstand helium (He) and hydrogen (H) fluxes from the plasma environment[3].

Tungsten (W) has been recognized as a leading plasma-facing material due to its remarkable properties. Its refractory nature, high melting point, excellent thermal conductivity, low erosion rates, and outstanding thermomechanical stability make it highly suitable for use as the first wall material in ITER and future fusion reactors[3–5]. However, the formation of helium (He) bubbles during helium irradiation raises significant concerns about the long-term structural integrity of plasma-facing components manufactured from pure tungsten[4,6]. In recent years, high-entropy alloys (HEAs) have been identified as promising alternatives to traditional materials for use in extreme environments[7–16].

Among these, W-based HEAs have demonstrated exceptional properties, including high melting points and superior mechanical performance at elevated temperatures, surpassing those of Ni-based superalloys and nanocrystalline tungsten[13,14]. Additionally, W-based HEAs have exhibited significantly enhanced radiation resistance compared to pure tungsten[7–9,11,12,15–17]. For instance, alloys such as MoNbTaVW and MoNbTaTiW have shown greater irradiation resistance than pure tungsten[7,12]. However, the inclusion of high-activation elements such as Mo and Nb in these alloys raises concerns regarding their long-term applicability as plasma-facing materials (PFMs). To address this issue, W-Ta-Cr-V HEAs have been developed as a potential solution[8,9,11,15,16]. An Experimental study on $W_{38}Ta_{36}Cr_{15}V_{11}$ revealed remarkable irradiation resistance when exposed to 1 MeV $Kr^{+2}$ ions, even at high irradiation doses of up to 8 displacements per atom (dpa) at 800° C[9]. Unlike pure tungsten, this alloy exhibited no evidence of dislocation loop formation under these conditions[9]. Furthermore, this HEA exhibited exceptional resistance to helium (He) bubble damage at 1223 K, with small (~2–3 nm) bubbles growing uniformly at a slow rate and no preferential formation on grain boundaries[15]. Molecular dynamics (MD) simulations further revealed that, although the number of Frenkel pairs (FPs) created in the primary damage state of $W_{38}Ta_{36}Cr_{15}V_{11}$ is higher than in pure tungsten, the interstitial cluster size and dislocation loop density are significantly lower[17]. These findings highlight the superior radiation resistance of this HEA[17]. Furthermore, ab initio simulations have attributed the radiation resistance of WTaCrV alloy to the slowed interstitial diffusion and the high recombination probability of interstitials and vacancies in the alloy[18]. An experimental study on equimolar WTaCrV HEA has demonstrated the formation of helium (He) bubbles with diameters below 1 nm during helium irradiation, indicating exceptional resistance to He irradiation-induced defect accumulation[11]. These results highlight the alloy's microstructural stability and resistance to irradiation hardening. Additionally, a recent atomistic simulation study demonstrated the exceptional resistance of equimolar WTaCrV HEA to surface modifications caused by energetic helium ions, further highlighting their potential for plasma-facing materials[8].

The majority of studies on HEAs have focused on their radiation response to heavy ion, neutron, or He irradiation. However, to evaluate the radiation response of materials under fusion reactor-relevant conditions—characterized by fast neutrons and gas production from transmutation reactions—a more complex irradiation environment is required. Atwani et al. addressed this challenge by mimicking such conditions through dual-beam irradiation with 1 MeV $Kr^+$ and 16 keV $He^+$ ions to test the radiation resistance of $W_{29.4}Ta_{42}Cr_{5.0}V_{16.1}Hf_{7.5}$ and $W_{31}Ta_{34}Cr_{5.0}V_{27}Hf_{3.0}$ at 1173 K[16]. Their findings revealed no dislocation loops, even after 8.5 displacements per atom (dpa) and 9.13% He implantation. However, cavities were observed under these conditions[16].

Despite these promising results, no fundamental studies have yet been conducted to investigate the formation and evolution of defects in W-Ta-Cr-V HEAs under complex irradiation environments. Specifically, the combined effects of pre-existing helium (introduced through He injection) followed by neutron or ion irradiation remain unexplored. In this work, we present a direct comparison of the radiation resistance of tungsten and equimolar WTaCrV HEA with 0%, 1%, and 2% He, added during the simulation setup. Displacement cascade overlap simulations were employed to mimic neutron irradiation. These simulations, widely used to model radiation damage in metals and alloys, provide a robust platform for studying the production and evolution of defects[12,19,20]. Utilizing a recently developed interatomic potential—constructed based on density functional theory (DFT) data for WTaCrV-He properties and validated against experimental observations of He bubble growth[8,10]— MD simulations provide a reliable platform for evaluating the comparative performance of these materials. The insights obtained from this study will enhance our understanding of the behavior of plasma-facing components (PFCs) under complex irradiation conditions. Furthermore, this work evaluates whether W-Ta-Cr-V HEAs remain strong candidates for advanced plasma-facing materials in the demanding environments of fusion reactors.

## Results

Figure 1 compares the evolution of defects in pure tungsten (W) and WTaCrV HEA in the absence of helium as a function of dose, up to 550 successive cascades, corresponding to 0.2 displacements per atom (dpa) based on the NRT formula at 300 K. The results indicate that, while the concentration of Frenkel pairs (FPs) is higher in the HEA compared to pure W (Fig. 1a), the tendency of interstitials to form clusters is significantly lower in the WTaCrV HEA (Fig. 1b, 1d, and 1e). In pure W, large interstitial clusters and dislocation loops are observed, whereas in the WTaCrV HEA, no dislocation loops or interstitial clusters larger than 10 atoms are detected. Notably, the largest interstitial cluster in pure W is nearly two orders of magnitude larger than the largest interstitial cluster in the WTaCrV alloy (Fig. 1b). For vacancies, the majority in both W and the HEA appear as single vacancies (Fig. 1f and 1g). Additionally, approximately 2% of the vacancies in both materials form vacancy clusters, typically ranging in size from 2 to 4 vacant sites. Larger vacancy clusters are observed in the HEA but represent only a small fraction of the total vacancy population.

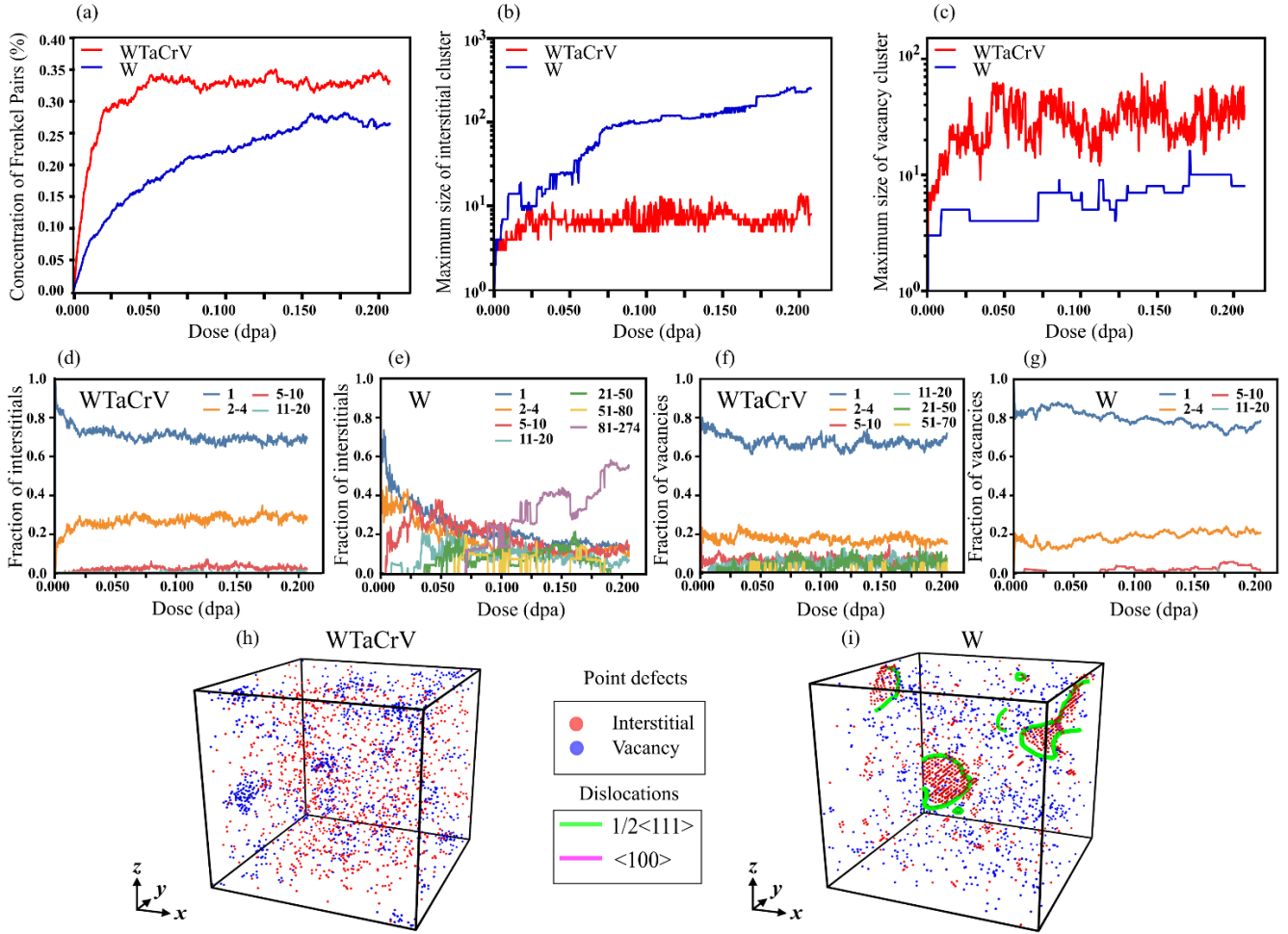

Fig. 1: Analysis and visualization of defect evolution in W and WTaCrV HEA under cascade overlap simulations: (a) Concentration of Frenkel pairs as a function of dose. (b) Maximum size of interstitial clusters as a function of dose. (c) Maximum size of vacancy clusters as a function of dose. Fraction of interstitials in different cluster size ranges as a function of dose in (d) W and (e) WTaCrV HEA. Fraction of vacancies in different cluster size ranges as a function of dose in (f) W and (g) WTaCrV HEA.

Visualization of defects at 0.2 dpa, including vacancies (blue), interstitials (red), and dislocations (green for ½<111> loops and pink for <100> loops) for (h) WTaCrV HEA and (i) W.

We introduced 1% helium atoms into the simulation box for each system and subsequently performed hundreds of cascade overlap simulations to reach 0.2 dpa. Figure 2 compares the evolution of defects in W and WTaCrV HEA with 1% helium atoms as a function of dose. The results indicate that the concentration of FPs is significantly higher in W compared to the HEA (Fig. 2a). Comparing this with Fig. 1a reveals that the introduction of 1% He dramatically increases the FP concentration in W at 0.2 dpa (approximately five times higher), while the FPs concentration in WTaCrV HEA increases by only 1.5 times compared to the case without He. Figures 2b, 2d, 2e and 2h show that in the HEA, no dislocation loops or large interstitial clusters are observed, whereas in W, a significant fraction of interstitials forms large dislocation loops. Comparing these results with Fig. 1 reveals that the introduction of 1% He notably increases the fraction of interstitials in large loops in W at 0.2 dpa. Figures 2c, 2f, and 2g provide insights into vacancy and He clusters. Figure 2c shows that the growth of the largest vacancy clusters is proportional to the growth of the largest He clusters, as vacancies are a primary source of He trapping. This leads to the formation of $V_n He_m$ bubbles, where $n$ represents the number of vacancies and $m$ represents the number of helium atoms. The presence of 1% He increases the fraction of vacancies in larger clusters for both W and the HEA, as shown by comparing Fig. 1f-g with Fig. 2f-g. In WTaCrV, vacancies predominantly form clusters of 11-20 and 20-51 vacant sites, whereas in W, the fraction of vacancies in 5-10 site clusters increases with dose. Figure 2h illustrates that in the HEA, large vacancy clusters are closely associated with He clusters, forming $V_n He_m$ bubbles. These bubbles are relatively well-separated, with surrounding regions containing single He atoms or very small clusters (2 – 4 sites), and exhibit spherical shapes. In contrast, Fig. 2i shows that in W, He bubbles are more inhomogeneous in size, shape, and distribution.

Figure 3 illustrates the radiation response of both systems under cascade simulations with 2% helium atoms introduced into the simulation box at the start of the simulations. Increasing helium concentration to 2% significantly impacts FP formation. At 0.2 dpa, FPs concentrations in W are 2.4 times higher than in the HEA (Fig. 3a). Comparing Fig. 2a with Fig. 3a shows that increasing the helium content from 1% to 2% raises the FP concentration in W by 2.4 times and in WTaCrV HEA by 2.6 times at 0.2 dpa. Notably, dislocation loops are absent in the HEA even at 0.2 dpa with 2% helium (Fig. 3h). However, the formation of larger interstitial clusters is observed in this system, with an increased fraction of clusters in the 5-20 atom size range (Fig. 3d). Additionally, a small fraction of interstitials forms a cluster containing 289 interstitials (Fig. 3b). By comparing Figures 2c, 2f, and 2h with Figures 3c, 3f, and 3h, it becomes clear that increasing helium content leads to the growth of $V_n He_m$ clusters in the HEA under irradiation. Figure 3f reveals that, in the HEA with 2% He and irradiated up to 0.2 dpa, the majority of vacancies are concentrated in clusters ranging from 21 to 50 vacancies. The growthing of $V_n He_m$ clusters with increasing helium content is also observed in W. Figure 3c shows that the largest vacancy clusters in W are comparable in size, or even slightly larger, than those in the HEA after 0.1 dpa. Additionally, Figure 3g indicates that the fractions of vacancies as single vacancies, 2-4 vacancy clusters, 5-10 vacancy clusters, and 11-20 vacancy clusters are nearly identical. However, a slightly smaller fraction is observed to accumulate in the 21-50 vacancy cluster size range.

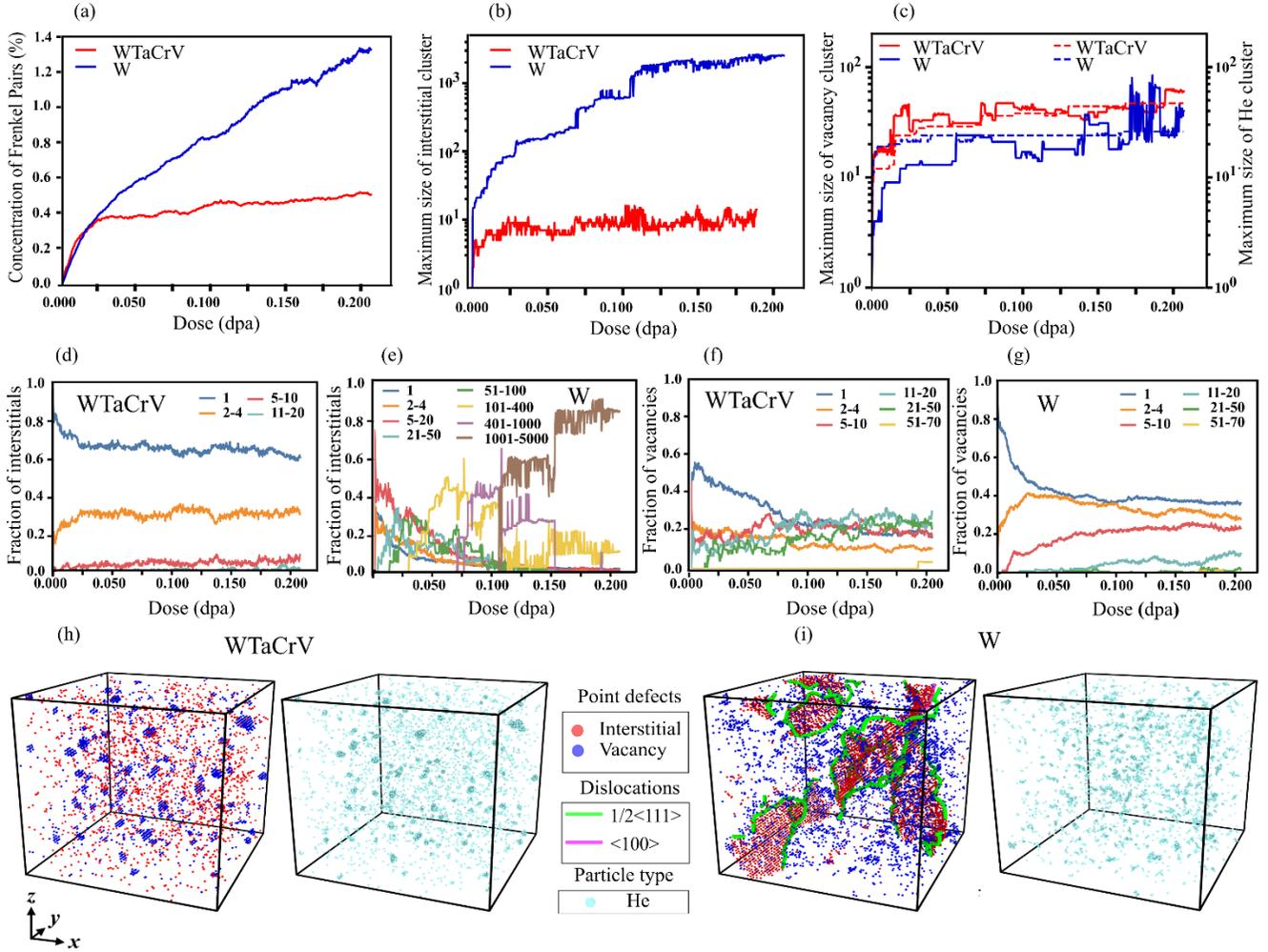

Fig. 2: Analysis and visualization of defect evolution in W and WTaCrV HEA with 1% He atoms inserted, under cascade overlap simulations: (a) Concentration of Frenkel pairs as a function of dose. (b) Maximum size of interstitial clusters as a function of dose. (c) Maximum size of vacancy clusters and He clusters as a function of dose. Fraction of interstitials in different cluster size ranges as a function of dose in (d) W and (e) WTaCrV HEA. Fraction of vacancies in different cluster size ranges as a function of dose in (f) W and (g) WTaCrV HEA. Visualization of defects at 0.2 dpa for (h) WTaCrV HEA and (i) W: one frame displays vacancies (blue), interstitials (red), and dislocations (green for ½<111> loops and pink for <100> loops), while another frame shows the positions of He atoms (light blue).

In W, the presence of 2% helium significantly promotes the formation of massive interstitial clusters during cascade overlap simulations. Figures 3b and 3e show the development of a large interstitial network with increasing dose, reaching a size of up to 12,869 atoms at 0.2 dpa and forming a distinct cross-like structure (Fig. 3i). The detailed progression of these interstitial clusters can be observed in Movie S1, while Figs. 4d-g provide two-dimensional snapshots showing the formation and evolution of this extensive polygonal interstitial network. Dislocation behavior in W is further analyzed in Fig. 4. As shown in Figs. 4a-c, increasing helium content leads to higher total dislocation densities. However, in W with 2% He, a noticeable drop in dislocation density begins at 0.05 dpa, ultimately leading to the formation of a large polygonal interstitial network by 0.19 dpa (Figs. 4d-g). Figures 4b-c indicate that both ½<111> and <100> dislocation loops form in W, with ½<111> dislocations being the predominant type in all cases.

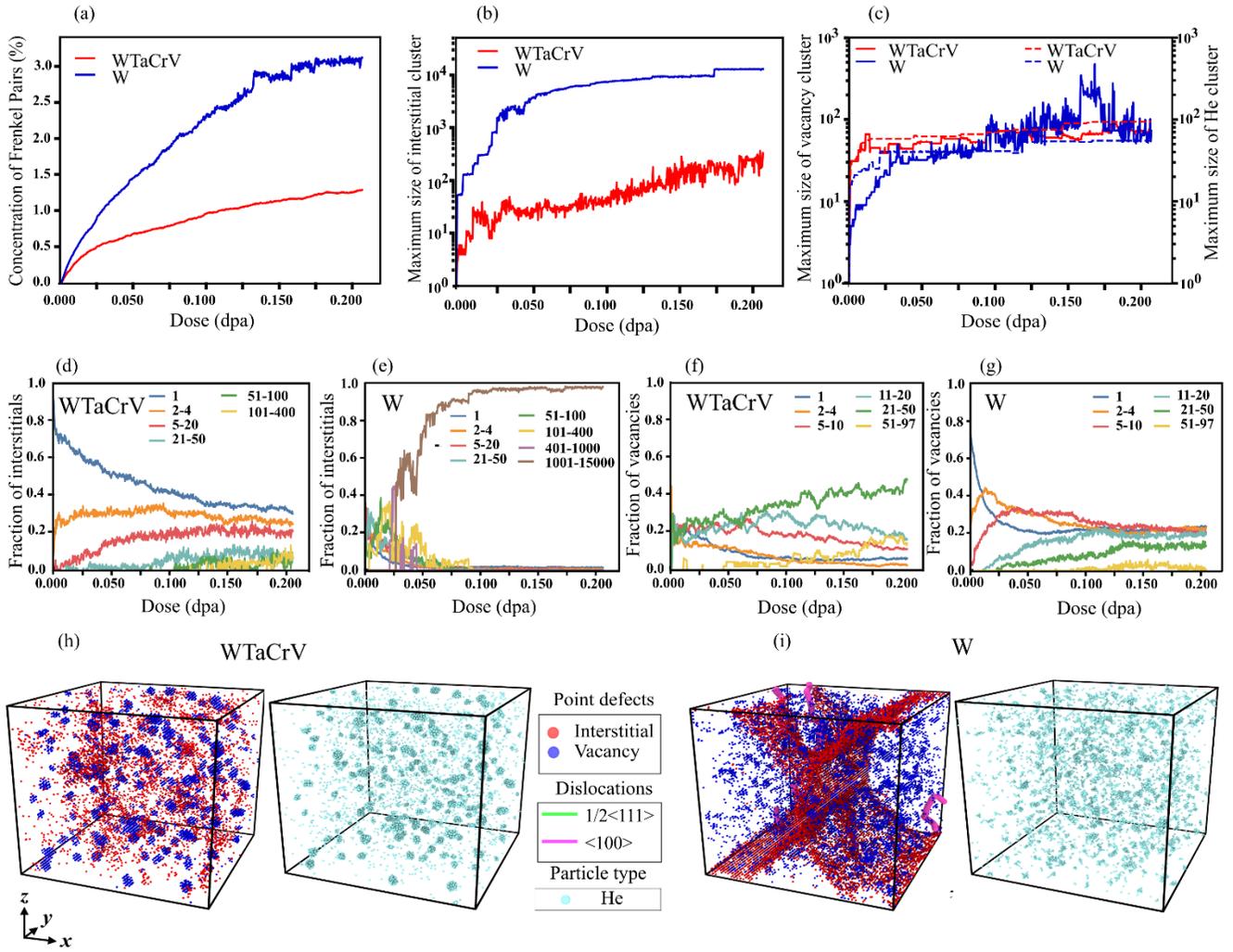

Fig. 3: Analysis and visualization of defect evolution in W and WTaCrV HEA with 2% He atoms inserted, under cascade overlap simulations: (a) Concentration of Frenkel pairs as a function of dose. (b) Maximum size of interstitial clusters as a function of dose. (c) Maximum size of vacancy clusters and He clusters as a function of dose. Fraction of interstitials in different cluster size ranges as a function of dose in (d) W and (e) WTaCrV HEA. Fraction of vacancies in different cluster size ranges as a function of dose in (f) W and (g) WTaCrV HEA. Visualization of defects at 0.2 dpa for (h) WTaCrV HEA and (i) W: one frame displays vacancies (blue), interstitials (red), and dislocations (green for ½<111> loops and pink for <100> loops), while another frame shows the positions of He atoms (light blue).

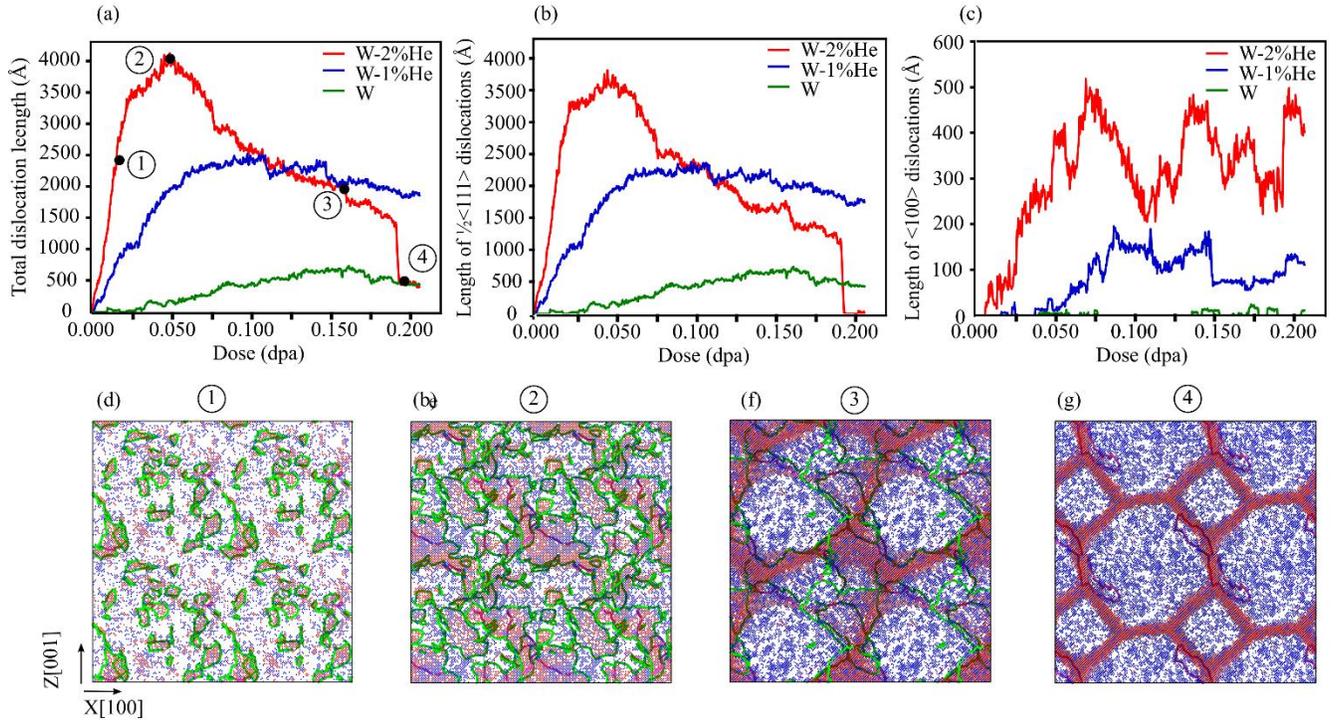

Fig. 4: Comparison of (a) total dislocation lengths, (b) ½<111> dislocation lengths, and (c) <100> dislocation lengths as a function of dose for tungsten with 0%, 1%, and 2% preexisting He. (d‒g) Two-dimensional frames illustrating the evolution of defects, including vacancies (blue), interstitials (red), and dislocations (green for ½<111> loops and pink for <100> loops), in W containing 2% He at four points indicated in (a). For better visualization, the simulation box is replicated twice in the x and z directions using periodic boundary conditions

## Discussion

Figure 1 demonstrates that the concentration of Frenkel pairs (FPs) in WTaCrV HEA is higher than in W up to 0.2 dpa. However, while interstitial clusters in W grow to significantly larger sizes compared to the HEA, vacancy clusters are slightly larger in the HEA. These results align with recent molecular dynamics (MD) simulations, which utilized a newly developed classical interatomic potential for WTaCrV HEA[17]. The primary radiation damage in W and $W_{38}Ta_{36}Cr_{15}V_{11}$ was compared under various primary knock-on atom (PKA) energies ranging from 1 to 100 keV. The simulations revealed a similar trend to what was observed in the cascade overlap simulations at the primary stage of damage[17]. Additionally, a recent MD study employing cascade overlap simulations and a machine learning-based interatomic potential compared defect evolution in W and WTaV alloy up to 0.4 dpa[12]. The results from that study are broadly consistent with ours, with one notable difference: in W, the concentration of FPs is lower than in WTaV up to ~0.1 dpa, whereas in our case, this trend persists up to 0.2 dpa. This discrepancy arises from differences in simulation size and the displacement energy ($E_d$) values used, which affect dose calculations in NRT simulations. Despite this, the saturation FP concentration in WTaV (~0.3‒0.35%) observed in that study aligns closely with the saturation observed for WTaCrV in our study[12]. Moreover, the maximum sizes of vacancy and interstitial clusters in WTaCrV from our simulations are comparable to those reported for WTaV. In contrast, in W, the maximum vacancy cluster size is smaller than in WTaV (at least up to 0.1 dpa), while the maximum interstitial cluster size in W is significantly larger than in WTaV, consistent with our observations[12].
7

When cascade overlap simulations are performed in W with pre-existing helium, the formation of interstitial clusters and the density of dislocation loops increase significantly (Figs. 1-4). Experimentally, the formation of dislocation loops and lines in W under dual-beam irradiation (involving He ion and heavy ion beams) has been observed, while the formation of bubbles strongly depends on helium content and temperature[21,22]. For instance, in an experimental study with 0.3% helium in the peak damage region under dual-beam irradiation ($He^+$ and $Fe^+$), bubbles were only observed at temperatures of 500°C and above[21]. In our simulations, increasing helium content enhances the tendency of vacancies to form larger clusters. The formation and growth of $V_n He_m$ clusters are expected to reduce interstitial-vacancy recombination and, consequently, lead to a higher interstitial concentration in W.

When comparing the evolution of defects in WTaCrV HEA with 0%, 1%, and 2% helium, as shown in Figures 1, 2, and 3, several observations become apparent. First, no dislocation loops are formed in any of these cases. This result is consistent with findings reported in the experimental literature. Specifically, it has been demonstrated that for $W_{38}Ta_{36}Cr_{15}V_{11}$ irradiated with heavy ions up to 8 dpa at high temperatures, no dislocation loops were observed when helium content was 0%[9]. Similarly, in another experimental study using dual-beam ion irradiation with $Kr^+$ and $He^+$ ions, no dislocation loops were observed in $W_{29.4}Ta_{42}Cr_{5.0}V_{16.1}Hf_{7.5}$ and $W_{31}Ta_{34}Cr_{5.0}V_{27}Hf_{3.0}$ HEAs, even after 8.5 dpa at elevated temperatures, despite the helium concentration reaching 9.13%[16]. The underlying mechanism for this behavior can be attributed to DFT calculations conducted by Zhao[18], which highlight the slowed interstitial diffusion in the WTaCrV alloy. This reduced mobility inhibits the clustering of interstitials necessary for dislocation loop formation.

The second observation is that the fraction of vacancies in large vacancy clusters, as well as helium clusters, increases with higher helium content. The growth of $V_n He_m$ clusters follows a distinct pattern: the clusters are well-spaced, with surrounding regions containing isolated He atoms or very small helium clusters (2-4 sites), and they exhibit a spherical morphology. In contrast, the bubble distribution morphology in W differs, featuring helium clusters of varying sizes that are distributed with similar populations. While WTaCrV HEA is resistant to the formation of large cavities under helium ion irradiation[11,15], cavity formation in this family of alloys has been reported under dual-beam irradiation at 1173 K[16]. Several mechanisms explain the growth of bubbles in WTaCrV HEA under cascade overlap simulations. First, helium atoms exhibit lower diffusivity in WTaCrV alloy compared to W. This reduced mobility is evident in Fig. 5, which shows the trajectory paths of a He atom in both systems. This low diffusivity is one of the reasons why experimentally, only small bubbles are observed in WTaCrV when irradiated solely with helium atoms.

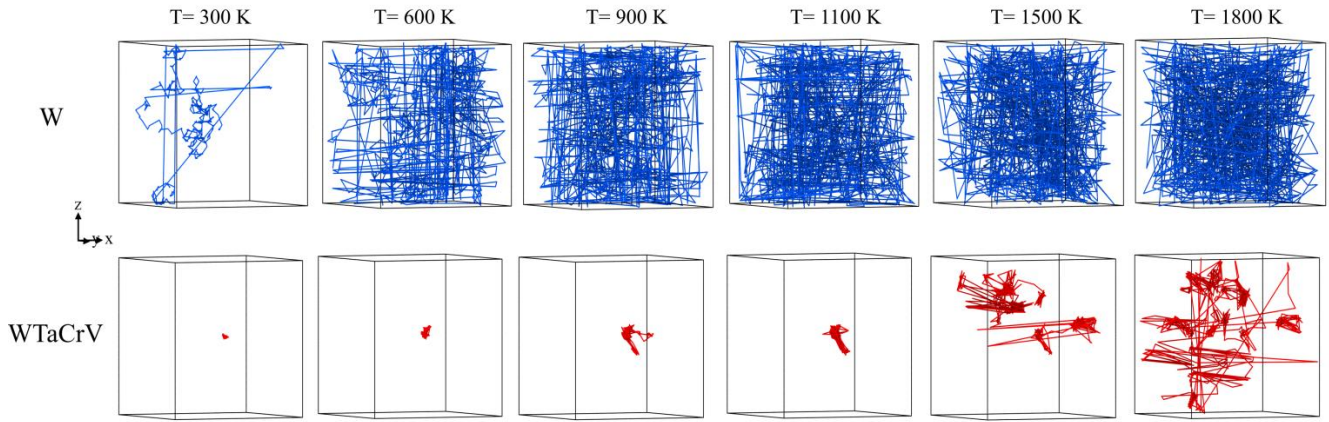

Fig. 5: Trajectory paths of a He atom in W (upper panel) and WTaCrV HEA (lower panel) at different temperatures. Simulations were conducted for up to 4 ns.

Another key factor is the dependence of He cluster formation on vacancy availability. In WTaCrV, the binding energy of vacancies to He clusters is relatively high, facilitating bubble growth. Studies have shown that the binding energy between helium atoms (in the absence of vacancies) is negligible in WTaCrV HEA, whereas in W, it becomes significant, particularly as the number of helium atoms in the cluster increases[8]. However, for $V_nHe_m$ clusters with $n>2$, the mean binding energy ranges from 1-2 eV, enhancing stability and promoting growth in the HEA[8,10]. Figure 6 demonstrates that in WTaCrV, even at 1800 K where He atoms can diffuse effectively, He clusters do not form without vacancies, whereas in W, He clusters can form even in the absence of vacancies. Additionally, Fig. 6 illustrates the formation of $V_nHe_m$ clusters in the presence of 9 vacancies at different temperatures for both systems. In WTaCrV HEA, formation of bubbles and growth under cascade overlap simulations can be described in three distinct phases. Initially, vacancies are created via displacement cascade mechanisms, and helium atoms diffuse into these vacancies to form $V_nHe_m$ bubbles. Helium migration in WTaCrV alloy is limited due to the high migration barrier of He atoms, and mobility is primarily induced by the high energy of recoil atoms and localized thermal spikes generated during displacement cascades. As the simulation progresses, bubble growth occurs through two primary mechanisms. First, single helium atoms are incorporated into existing $V_nHe_m$ clusters, facilitated by vacancy trapping sites. Helium atoms can displace lattice atoms during this process, increasing the number of vacancies in the cluster. Second, nearby $V_nHe_m$ bubbles can coalesce under the influence of energy from displacement cascades, resulting in larger bubbles with higher helium content (Fig. 7a). Finally, the stability of large $V_nHe_m$ bubbles becomes apparent. Due to their size and helium content, these bubbles are energetically favorable and highly resistant to deformation or shrinkage under further displacement cascades. This stability, as shown in Fig. 7b, is critical for understanding bubble evolution under irradiation conditions.

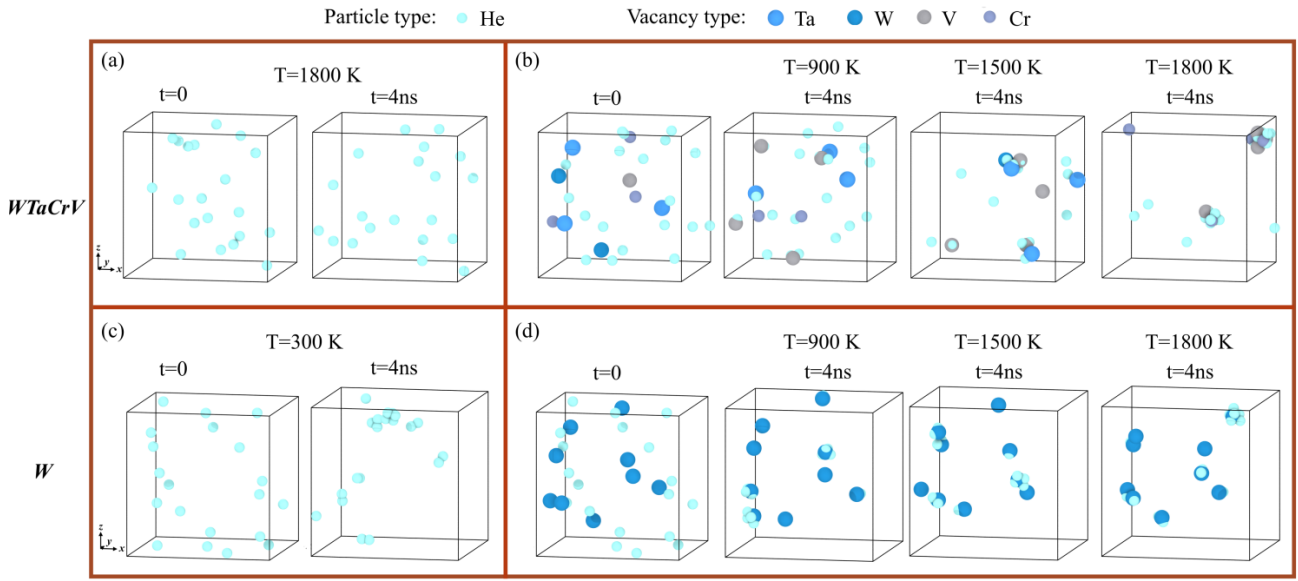

Fig. 6: Evolution of 20 He atoms (1%) at different temperatures in WTaCrV HEA and W: (a) WTaCrV HEA without preexisting vacancies, (b) WTaCrV HEA with 9 preexisting vacancies, (c) W without preexisting vacancies, and (d) W with 9 preexisting vacancies.

In summary, we investigated defect evolution in W and WTaCrV HEA with and without preexisting helium atoms under cascade overlap simulations up to 0.2 dpa at 300 K using molecular dynamics. In W, we observed the formation of dislocation loops and large interstitial clusters, regardless of preexisting helium. Increasing helium content and dose resulted in higher dislocation densities and the development of a polygonal interstitial network. The size and quantity of helium clusters grew with increasing dose and helium content, with notable differences in their size and spatial distribution between W and WTaCrV alloy. While WTaCrV HEA exhibited strong resistance to dislocation loop formation and large interstitial clusters, it was less resistant to bubble formation, particularly at higher helium concentrations. These findings suggest the necessity of revisiting the design principles for HEAs to achieve better radiation resistance under such complex irradiation environments.

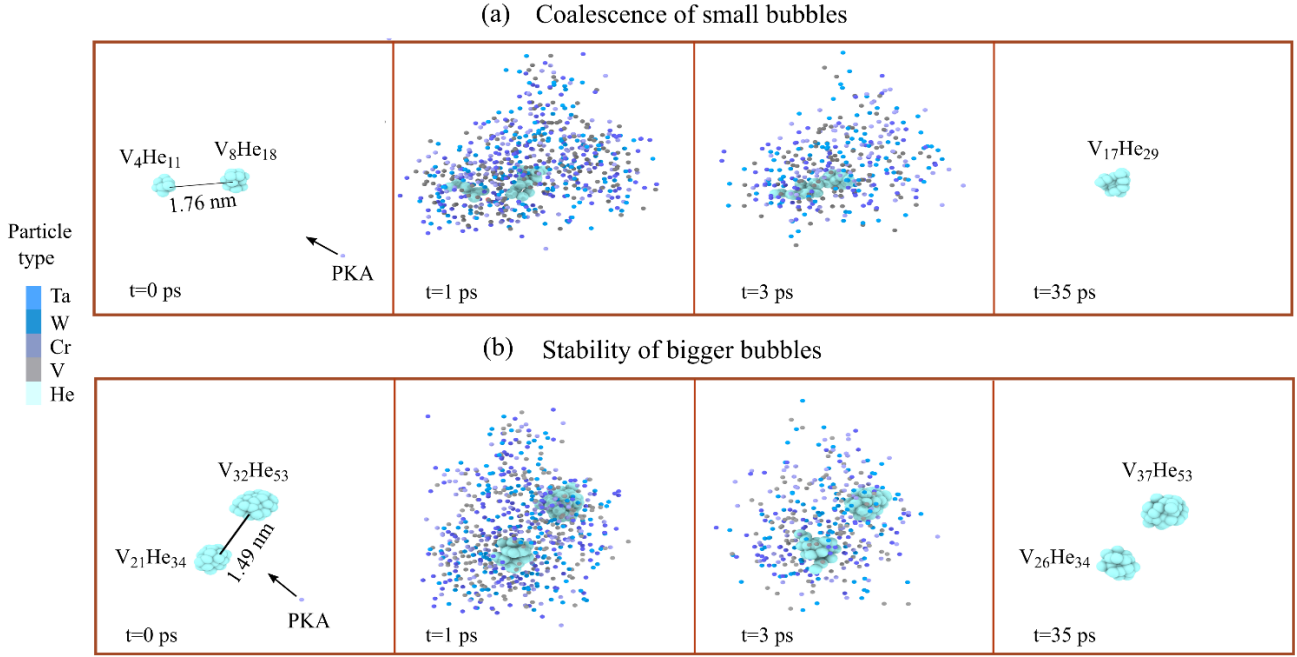

Fig. 7: (a) Snapshots illustrating how displacement cascade overlap induces the coalescence of two small $V_nHe_m$ bubbles. (b) Snapshots demonstrating the stability of a large bubble under displacement cascades. To visualize the displacement cascade area, only metallic atoms with kinetic energies exceeding 0.87 eV are shown.

**Methods**

All molecular dynamics simulations in this study were conducted using the Large-scale Atomic/Molecular Massively Parallel Simulator (LAMMPS) [23]. We employed the embedded atom method (EAM) potential developed by Zhou et al. to describe interatomic forces between metallic atoms[8,24]. The interactions among He atoms were modeled using the pair potential developed by Becker[25], while the He-metal interactions were based on the pair potential formalism introduced by Juslin and Nordlund (JN)[8,26,27]. To accurately capture short-range interactions, all potentials were smoothly integrated with the Ziegler-Biersack-Littmark (ZBL) repulsive potential[8,28]. We created multiple simulation boxes for cascade overlap simulations in W and WTaCrV systems, with and without preexisting He. Each box consisted of 55×55×55 bcc unit cells aligned along the <100>, <010>, and <001> directions, containing 332,750 W atoms. HEA configurations were generated by randomly replacing W atoms with Ta, Cr, and V atoms to achieve an equimolar composition (W:Ta:Cr:V = 1:1:1:1). Helium atoms were then added randomly at 1% and 2% concentrations relative to the total number of atoms for He-containing systems. Periodic boundary conditions were applied in all directions, and energy minimization was performed using the conjugate gradient (CG) method to stabilize the system. The system was then equilibrated at 300 K and zero pressure using the NPT ensemble for 100 ps. To simulate irradiation, a central metallic atom was assigned recoil energy of 10 keV in a random direction, initiating a collision cascade. The system was maintained at a constant volume, with the borders (outer 5 Å of the simulation box) controlled at 300 K using a Berendsen thermostat [29] for the 20 ps. Subsequently, the entire box was stabilized at constant pressure using a Berendsen barostat for an additional 10 ps. This irradiation process was repeated 550 times per simulation cell, achieving ~0.2 dpa based on the NRT approximation [30] i.e. $\left(\frac{n}{N}\right) \times \frac{E_{recoil}}{2E_d}$, where n indicates the number of cascades, N represents the total number of atoms in the box, $E_{recoil}$ refers to the energy of recoiled atom, and $E_d$ signifies the displacement threshold energy. Ed= 40 eV was chosen for all the metallic elements in accordance with Ref.[9]. To achieve uniform irradiation, the simulation box was randomly shifted across periodic boundaries following each cascade. Vacancy and interstitial concentrations were assessed using the Wigner-Seitz method[31,32], while cluster distributions were identified based on atomic nearest-neighbor distances. Dislocation structures were analyzed using the Dislocation Extraction Algorithm (DXA)[33], with all analyses conducted in OVITO software[34]. For the simulations corresponding to Fig. 5 and Fig. 6, smaller simulation boxes containing approximately 2000 atoms were created, with dimensions of 30.764 Å × 30.764 Å × 30.764 Å along the <100>, <010>, and <001> directions. The same procedures used for generating the WTaCrV HEA composition, as well as the energy minimization and equilibration steps applied to the larger samples, were employed for these smaller simulation boxes.

**Acknowledgments**

This research was funded by the European Union Horizon 2020 research and innovation program under NOMATEN Teaming grant (agreement no. 857470) and from the European Regional Development Fund via the Foundation for Polish Science International Research Agenda PLUS program grant No. MAB PLUS/2018/8. The publication was created within the framework of the project of the Minister of Science and Higher Education "Support for the activities of Centres of Excellence established in Poland under Horizon 2020″ under contract no. MEiN/2023/DIR/3795. We acknowledge the computational resources provided by the High Performance Cluster at the National Centre for Nuclear Research in Poland.

**Competing interests**
The authors declare no competing interests
.
**Data availability**
Data will be made available on request
.